\documentclass[sigconf]{acmart}
\usepackage{multirow}
\usepackage{subcaption}
\usepackage{enumitem}
\AtBeginDocument{%
  }
\usepackage{amsmath}
\usepackage{mathtools}
\usepackage{algorithm}
\usepackage{algpseudocode}

\usepackage{array}
\usepackage{graphicx}
\usepackage{booktabs}
\usepackage{dsfont}
\usepackage{balance}

\setcopyright{acmlicensed}
\copyrightyear{2024}
\acmYear{2024}
\acmDOI{10.1145/3627673.3679986}

\acmConference[CIKM '24]{33rd ACM International Conference on
Information and Knowledge Management}{October 21--25,
  2024}{Boise, ID, USA}
 
\acmPrice{15.00}
\acmISBN{979-8-4007-0436-9/24/10}

\begin{document}

% \title{Reduced Cross-Entropy Loss for Sequential Recommendations with Large Item Catalogs}
\title{RECE: Reduced Cross-Entropy Loss for Large-Catalogue Sequential Recommenders}
% \title{Autoregressive generation with GPT-2 architecture for long-term sequential recommendations}


\author{Danil Gusak}    
\orcid{0009-0008-1238-6533}
\affiliation{%
  \institution{Skolkovo Institute of Science and Technology}
  \country{}
}
\affiliation{%
  \institution{HSE University}
  \city{Moscow}
  \country{Russian Federation}
}
\email{danil.gusak@skoltech.ru}
\authornote{Both authors contributed equally to the paper}

\author{Gleb Mezentsev}
\orcid{0009-0003-7591-3082}
\affiliation{%
  \institution{Skolkovo Institute of Science and Technology}
  \city{Moscow}
  \country{Russian Federation}
}
\email{gleb.mezentsev@skoltech.ru}
\authornotemark[1]

\author{Ivan Oseledets}
\orcid{0000-0003-2071-2163}
\affiliation{%
  \institution{Artificial Intelligence Research Institute}
  \country{}
}
\affiliation{%
  \institution{Skolkovo Institute of Science and Technology}
  \city{Moscow}
  \country{Russian Federation}
}
\email{oseledets@airi.net}

\author{Evgeny Frolov}
\orcid{0000-0003-3679-5311}
\affiliation{%
  \institution{Artificial Intelligence Research Institute}
  \country{}
}
\affiliation{%
  \institution{Skolkovo Institute of Science and Technology}
  \country{}
}
\affiliation{%
  \institution{HSE University}
  \city{Moscow}
  \country{Russian Federation}
}
\email{frolov@airi.net}

% \renewcommand{\shortauthors}{Gusak, et al.}


\begin{abstract}
  Scalability is a major challenge in modern recommender systems. In sequential recommendations, full Cross-Entropy (CE) loss achieves state-of-the-art recommendation quality but consumes excessive GPU memory with large item catalogs, limiting its practicality.

Using a GPU-efficient locality-sensitive hashing-like algorithm for approximating large tensor of logits, this paper introduces a novel RECE (\textbf{RE}duced \textbf{C}ross-\textbf{E}ntropy) loss. RECE significantly reduces memory consumption while allowing one to enjoy the state-of-the-art performance of full CE loss. Experimental results on various datasets show that RECE cuts training peak memory usage by up to $12$ times compared to existing methods while retaining or exceeding performance metrics of CE loss.
The approach also opens up new possibilities for large-scale applications in other domains.

\end{abstract}

\begin{teaserfigure}
\vspace{-15pt}
\setlength{\abovecaptionskip}{4pt}
\centering
    \includegraphics[width=1.0\textwidth]{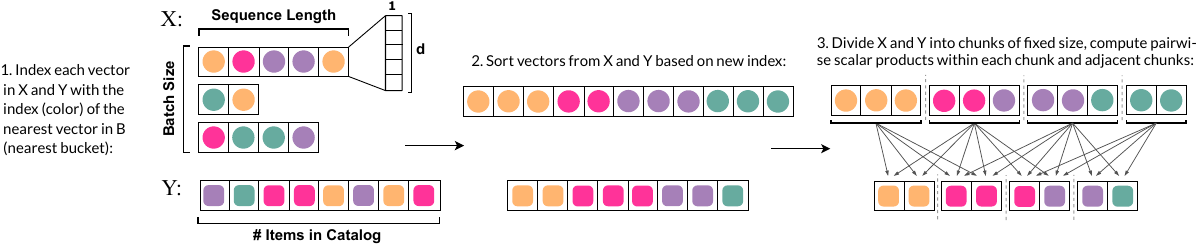}
    \caption{Simplified example illustrating the bucketing, sorting, chunking, and scalar product calculation steps of the proposed RECE approach. Different colors correspond to belonging to different buckets.}
    \label{fig:part_alg}
\end{teaserfigure}





\maketitle

\section{Introduction}
\label{sec:intro}
\newlist{customitemize}{enumerate}{1}
\setlist[customitemize,1]{label=(\arabic*), leftmargin=*}

\begin{figure}[]
\setlength{\abovecaptionskip}{6pt}
\setlength{\belowcaptionskip}{-12pt}
    \centering
    \includegraphics[width=1\columnwidth]{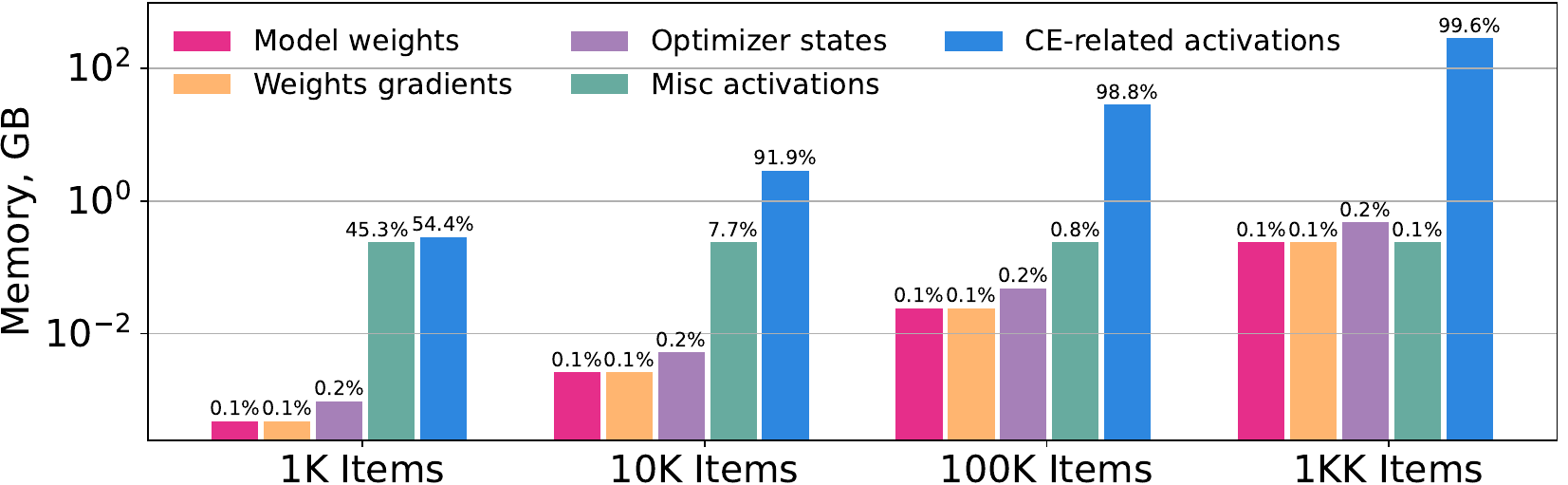}
    \caption{Impact of different components on peak GPU memory usage during SASRec training with Cross-Entropy loss. Measurements were conducted using PyTorch profiling tools.}
    \label{fig:memory}
    \Description[]{}
\end{figure}

In collaborative filtering, recent state-of-the-art models increasingly adopt sequential approaches to predict the next item a user might choose based on past activity. By considering the sequence of interactions, such systems make timely and relevant recommendations, like suggesting phone accessories following a phone purchase.

Transformer architectures \cite{vaswani2017attention}, originally from natural language processing (NLP), have been successfully adapted for \emph{next item prediction task} in recommender systems. Notable models include SASRec \cite{kang2018self} and BERT4Rec \cite{sun2019bert4rec}, inspired by GPT \cite{radford2018gpt} and BERT \cite{devlin2018bert} architectures. Initially, SASRec employed Binary Cross-Entropy (BCE) loss for training, but subsequent research showed that full Cross-Entropy (CE) loss \eqref{eq:ce} enables SOTA performance \cite{Klenitskiy_2023, Petrov_2023}, highlighting CE effectiveness.
CE loss, however, is memory-intensive due to the need to compute and store a large tensor of logits, making CE less scalable for larger item catalogs (Fig. \ref{fig:memory}). The challenge is \emph{to develop a reduced replacement for CE loss that maintains accuracy while operating within memory constraints similar to those of BCE}.

We propose a novel RECE (\textbf{RE}duced \textbf{C}ross-\textbf{E}ntropy) loss, \emph{which uses a selective computation strategy} to prioritize the most informative elements from input sequences and item catalog - those most likely to cause misclassifications. By approximating the softmax distribution over these elements, pre-identified using a GPU-friendly approximate search for maximum inner products, RECE \emph{eliminates the need to compute and store the full logit tensor} while mitigating inefficiencies of less selective negative sampling. We evaluate RECE integrated into SASRec on four datasets. This approach can also benefit other domains like NLP and search systems.

To summarize, the main contributions of this paper are:
\begin{customitemize}
  \item We propose RECE -- a memory-efficient approximation of CE loss, potentially applicable beyond sequential recommenders;
  \item We use RECE to train SASRec, conduct an extensive evaluation, and show that RECE significantly reduces the peak training memory without compromising the model performance.

\end{customitemize}

\section{Related Work}
\label{sec:literature}
Since the introduction of the Transformer architecture \cite{vaswani2017attention}, Transformer -based models have outperformed other approaches in sequential recommendations \cite{sun2019bert4rec, Xie2022ContrastiveLF, du2022contrastive}. SASRec \cite{kang2018self} shows state-of-the-art performance with CE loss \cite{Klenitskiy_2023}. However, large item catalogs in real-world applications require negative sampling methods or CE approximations to train such models. 

Uniform random sampling of negatives is a straightforward method
\cite{tang2018personalized, kang2018self}. It can be improved by increasing the number of negative samples, modifying BCE (\ref{eq:bce_2}) or CE (\ref{eq:ce_2}) \cite{Klenitskiy_2023} loss functions:
\begin{equation}
    \mathcal{L}_{BCE^+} = - \log (\sigma(logit_{i, +})) - \sum_{j \in I_n^-} \log (1 - \sigma(logit_{i, j})),
    \label{eq:bce_2}
\end{equation}
\begin{equation}
    \mathcal{L}_{CE^-} = -\log \frac{\exp(logit_{i, +})}{ \exp(logit_{i, +}) + \sum_{j \in I_n^-} \exp(logit_{i, j})},
    \label{eq:ce_2}
\end{equation}
where $I_n^-$ is a set of $n$ sampled negatives.

However, this method often lacks \emph{hard negatives} (negatives that the model misclassifies as positives), leading to overconfidence \cite{Petrov_2023}. Calibrating predicted scores can mitigate this  \cite{Petrov_2023}, resulting in SOTA performance, but leaves samples uninformative, suggesting possible improvement. Popularity-based sampling \cite{lian2020personalized, chen2022generating} is another approach, often better than uniform sampling but still outperformed by methods targeting hard negatives directly \cite{chen2022generating, rendle2014improving}. In-batch negative sampling \cite{hidasi2015session, hidasi2018recurrent} uses true class labels from other items in the batch, leveraging item popularity. More informative sampling methods approximate softmax distributions using matrix factorizations \cite{rendle2014improving}, adaptive n-grams \cite{bengio2008adaptive}, and kernel methods \cite{blanc2018adaptive, rawat2019sampled}. Two-step procedures select items with larger logits for loss computation \cite{bai2017tapas, chen2022generating, wilm2023scaling}. Methods targeting hard negatives include accumulating hard negatives for each user \cite{wang2021cross, ding2020simplify}, but this introduces memory overhead. Instead, hard negatives can be selected at each step using (approximate) maximum inner product search (MIPS) or nearest neighbor search (NNS) \cite{vijayanarasimhan2014deep, spring2017new, guo2016quantization, yen2018loss, lian2020personalized}. However, the implementations of these methods are not designed for GPU usage and are not easily adaptable due to their reliance on GPU-inefficient operations (like maintaining a hash table).

In summary, existing methods fail to target hard negatives effectively or are inefficient for GPU computations, resulting in suboptimal model performance. Our approach addresses this by ensuring an efficient search for hard negatives and batch processing compatibility, which improves GPU utilization.

\section{Reduced Cross-Entropy}
\label{sec:approach}
In this section, we propose a novel scalable approach for CE loss approximation that reduces memory requirements while maintaining performance, and discuss its wide applicability across domains.

Inspired by the studies \cite{kitaev2020reformer} for efficient attention approximation, our method utilizes locality-sensitive hashing for angular distance \cite{andoni2015practical} for the calculation of CE loss over the part of catalog that most affects gradient updates, finding this part in a GPU-friendly manner.

If we are predicting the next item $z_{i+1}$ (catalog index) for item $z_i$, with $x_i$ as the transformer's output for $z_i$, $y_j$ as the embedding of item $j$ (both of dimension $d$) and the model output score $logit_{i, j} = x_i^Ty_j$, then for catalog size $C$, the CE loss for item $z_i$:
\begin{equation}
    \mathcal{L}_{CE} = -\log(\text{softmax}(logit_i)_{z_{i+1}})\
    = -\log \frac{\exp(logit_{i, z_{i+1}})}{\sum_{c=1}^C \exp(logit_{i, c})}
    \label{eq:ce}
\end{equation}
\begin{equation}
    \frac{\partial \mathcal{L}_{CE}}{\partial logit_{i,k}} = \text{softmax}(logit_i)_k - \mathds{1}[k = z_{i+1}] \in \left(-1,1\right)
    \label{eq:ce_grad}
\end{equation}
The gradient of the CE loss with respect to logits ranges from $-1$ to $1$ (Eq. \ref{eq:ce_grad}). It is close to $1$ for high predicted probabilities of incorrect classes and close to $-1$ for low predicted probabilities of correct classes. We aim to compute only logits with the largest absolute gradient values to preserve the most information, identifying these cases in advance. While the correct class logit is known, finding large logits for negative classes is a harder task. We simplify this by searching for all large logits, essentially solving a MIPS problem.

To address this task, we propose the RECE approach, presented in Algorithm \ref{alg:rce_new} with Lines 3-12 depicted in Fig. \ref{fig:part_alg}. It starts by generating a set $B$ of $n_b$ random vectors (Line 2), then indexing of the transformer outputs $X$ (dimensions for batch size $s$ and sequence length $l$ are collapsed) and catalog item embeddings $Y$ with the index of the nearest vector from $B$ (Lines 3-4). We want to divide items from $X$ and $Y$ into groups based on these indices and to calculate logits only within groups. The idea is that two vectors sharing the nearest vector (in terms of dot product) are likely close to each other. The sizes of these groups could be different, and to perform later computations efficiently, we sort elements based on new indices and divide them into $n_c$ equal-sized chunks (Lines 5-11). The number of chunks $n_c$ can be selected larger than $n_b$ so that relevant items fall into the same chunk with a higher probability. For the same reason, for logit calculation, we also select items from the neighboring chunks (Line 12). Finally, we calculate logits for negative classes within chunks, compute positive logits ($\hat{Z}$ -- correct predictions matrix), determine the value of the loss function for each chunk, and average these values across chunks (Lines 12-16). For better performance, the described procedure (Lines 2-12) can be repeated in parallel over several rounds $r$. In this case, the value of loss function is calculated over an enriched set of negative examples. Duplicate item pairs are accounted for by subtracting from the calculated logit value the natural logarithm of the number of times the logit between these items was calculated over all rounds. For the experiments, we chose the optimal, in terms of peak memory, number of random vectors ($n_b^* = \sqrt{4 \alpha_{bc} (1 + 2 n_{ec}) \cdot \min(C, s\cdot l)}$, where  $\alpha_{bc} = n_b / n_c$, $n_{ec}$ is the number of neighboring chunks we look into). The memory complexity of our algorithm is then $2 r\sqrt{\alpha_{bc} (1+2 n_{ec}) \cdot \min(C, s\cdot l)} \cdot \max(C, s\cdot l)$. This is $\sqrt{\min(C, s\cdot l)} / (2r \sqrt{\alpha_{bc} (1+2 n_{ec})})$ times smaller than the memory size required for the full Cross-Entropy loss. The extended derivation is available in our GitHub repository\footref{github}.

\begin{algorithm}
   \caption{Reduced Cross-Entropy Loss}
   \label{alg:rce_new}
\begin{algorithmic}[1]
\small
   \State {\bfseries Input:} $\hat{Z} \in \mathbb{N} ^{s \cdot l}$, $X \in \mathbb{R} ^{s \cdot l \times d}$, $Y \in \mathbb{R} ^{C \times d}$, $n_b$, $n_c$
   \State $B = \text{randn}_{\scriptstyle{\mathcal{N}(0, 1)}}(n_b, d) = \{b_k\}_{k=1}^{n_b}$
   
    \State $I = \text{argmax}_b \text{ }  B X^\top$ \footnotesize\Comment{index of nearest $b_k$}\small

    \State $J = \text{argmax}_b \text{ }  B Y^\top$ 
    % \footnotesize\Comment{select top-$b_x$}\small

    \State $X = X[\text{argsort } I]$ \footnotesize\Comment{sorting based on new index}\small

    \State $Y = Y[\text{argsort } J]$ 
    % \footnotesize\Comment{select top-$b_x$}\small
    
    \State
    {\bfseries for $c$ in range($n_c$):} \footnotesize\Comment{divide into $n_c$ chunks}\small

    \State\hspace{\algorithmicindent} $I_c = \{i: c |X| / n_c \leq i < (c+1) |X| / n_c\}$ 
    
    \State\hspace{\algorithmicindent} $J_c = \{j: c |Y| / n_c 
    \leq j < (c+1) |Y| / n_c\}$ 
    % \footnotesize\Comment{select top-$b_y$}\small
   
   \State\hspace{\algorithmicindent} $X_{c} = X[I_{c}]$
   \State\hspace{\algorithmicindent} $Y_{c*} = Y[J_{c-1}\cup J_{c}\cup J_{c+1}]$ 
   \footnotesize\Comment{current and adjacent chunks}\small

   \State\hspace{\algorithmicindent} $logits^-_c = X_c {Y^\top}_{c*}  \in \mathbb{R} ^{|X|/n_c\times 3|Y|/n_c}$ \footnotesize\Comment{logits for wrong classes }\small

    \State\hspace{\algorithmicindent}
    {\bfseries for $i$ in range($ |X| / n_c$):} 
    \State\hspace{\algorithmicindent}\hspace{\algorithmicindent} $logit^+_{c, i} = \sum_{k=1}^{d} X[I_{c, i}]_k \cdot Y[\hat{Z}[I_{c, i}]]_k$  \footnotesize\Comment{logit for correct class}\small
    \State\hspace{\algorithmicindent}\hspace{\algorithmicindent} $loss_{c,i} = -\log \frac{\exp(logit^+_{c, i})}{\exp(logit^+_{c, i})+\sum_{j=1}^{3|Y|/n_c}\exp(logits^-_{c, i, j})}$
   
   \State $\mathcal{L}_{\text{RECE}} = \frac{1}{|X|} \sum_{c,i} loss_{c,i}$ \footnotesize\Comment{accumulation of loss values}\small
\end{algorithmic}
\end{algorithm}

% \subsection{Method Applicability}\label{sec:versatility}

In this work, we utilize SASRec \cite{kang2018self} as our base model due to its widespread use in the literature \cite{cls4rec, Petrov_2023} and its state-of-the-art performance in sequential recommendations with full Cross-Entropy loss \cite{Klenitskiy_2023}. Although our primary focus on the RECE method is on recommender systems, where managing large catalogs is a common challenge, the applicability of this approach extends to various domains such as NLP, search systems, computer vision tasks, bioinformatics, and other areas, where tasks with extensive vocabularies or large number of classes are a common bottleneck.

\section{Experimental Settings}
\label{sec:experiments}
\label{sec:experimental_settings}

\paragraph{Datasets}\label{sec:datasets}

We conduct our main experiments on four diverse real-world datasets: BeerAdvocate \cite{mcauley2012beer}, Behance \cite{behance}, Amazon Kindle Store \cite{kindle}, and Gowalla \cite{cho2011gowalla}. In line with previous research \cite{tang2018personalized, kang2018self, sun2019bert4rec}, we interpret the presence of a review or rating as implicit feedback. Additionally, following common practice \cite{kang2018self, rendle2010factorizing, zhang2019feature} and to ensure the number of items in datasets allows for computing full Cross-Entropy loss within GPU memory constraints, we exclude unpopular items with fewer than $5$ interactions and remove users who have fewer than $20$ interactions. The final dataset statistics are summarized in Table \ref{tab:datasetStats}. The number of items ranges from $22,307$ in BeerAdvocate to $173,511$ in Gowalla, allowing us to evaluate methods under various memory consumption conditions.

\begin{table}[]
\setlength{\abovecaptionskip}{3pt}
\footnotesize
\caption{\textbf{{Statistics of the datasets after preprocessing.}}} \label{tab:datasetStats}
\resizebox{1\columnwidth}{!}{%

\begin{tabular}{llrrrr}
\hline
Dataset                                              & Domain    & Users  & Items   & Interactions & Density \\ \hline
BeerAdvocate \cite{mcauley2012beer} & Food      & 7,606  & 22,307  & 1,409,494    & 0.83\%  \\
Behance \cite{behance}              & Art       & 8,097  & 32,434  & 546,284      & 0.21\%  \\
Kindle Store \cite{kindle}                & E-com     & 23,684 & 96,830  & 1,256,065    & 0.05\%  \\
Gowalla \cite{cho2011gowalla}       & Soc. Net. & 27,516 & 173,511 & 2,627,663    & 0.06\%  \\ \hline
\end{tabular}%
}
\end{table}

\paragraph{Evaluation}\label{sec:evals}

\begin{figure}[h]
\setlength{\abovecaptionskip}{2pt}
    \centering
    \includegraphics[width=1\columnwidth]{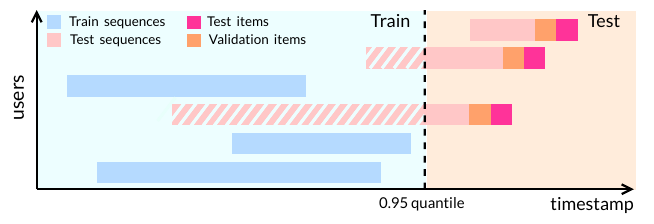}
    \caption{Temporal data splitting strategy.}
    \label{fig:split}
    \Description[]{}
\end{figure}

In offline testing, data splitting using the leave-one-out approach, often by selecting each user's last interaction, is common in previous studies \cite{Petrov_2023, Klenitskiy_2023}. However, this method can lead to data leakage affecting evaluation accuracy \cite{Ji_2023, meng2020exploring}. To mitigate this, we set a global timestamp at the $0.95$ quantile of all interactions \cite{frolov2022tensorbased}. Interactions before this timestamp are used for training, while interactions after -- for testing, keeping test users separate from the training data (Fig. \ref{fig:split}). For test users, we use their last interaction to evaluate model performance. This temporal split prevents "recommendations from future" bias \cite{meng2020exploring}, ensuring the model remains unaware of future interactions. In addition, we use the second-to-last interaction of each test user for validation to tune the model and to control its convergence via early stopping.

Following best practices \cite{Dallmann_2021, rocio21}, we use unsampled top-K ranking metrics: Normalized Discounted Cumulative Gain (NDCG@K) and Hit Rate (HR@K), with K $= 1, 5, 10$. Our goal is to balance time consumption, memory efficiency, and ranking performance, so we also measure training time and peak GPU memory during training.

\paragraph{Model and Baselines}\label{sec:baselines}

In our experiments, we use SASRec as the base model and enhance it with the proposed RECE loss. We focus on comparing \textbf{SASRec-RECE} with the model incorporating Binary Cross-Entropy loss with multiple negative samples (\textbf{SASRec-BCE$^+$}), detailed in Eq. \eqref{eq:bce_2}, and with SASRec employing full Cross-Entropy loss (\textbf{SASRec-CE}). Furthermore, we explore recent SOTA sampling-based variations of the loss function for SASRec proposed by Klenitskiy et al. \cite{Klenitskiy_2023} and Petrov et al. \cite{Petrov_2023}, denoted \textbf{SASRec-CE$^{-}$} and \textbf{gSASRec} (gBCE loss) respectively. All models are based on the adapted PyTorch implementation\footnote{\url{https://github.com/pmixer/SASRec.pytorch}} of the original SASRec architecture, and augmented with loss functions and sampling strategies from the respective papers. All code is available in our repository\footnote{\url{https://github.com/dalibra/RECE}\label{github}}.
\section{Results}\label{sec:results0}

\begin{figure*}[]
    \centering
    \setlength{\abovecaptionskip}{6pt} 
    \includegraphics[width=1.0\textwidth]{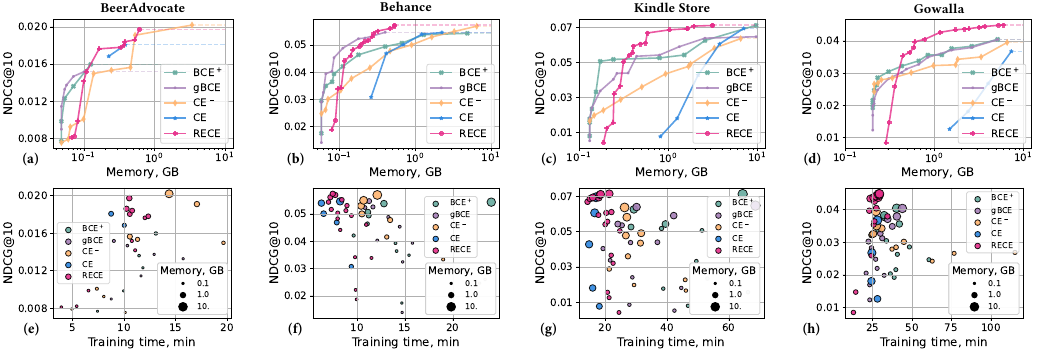}
    \caption{(a)-(d) Pareto front curves illustrating NDCG@10 for different memory budgets. (e)-(h) Same points plotted on NDCG@10 vs. Training time axes, with point sizes representing memory values from corresponding (a)-(d) plots.}
\label{fig:baselines}
\end{figure*}

Our experiments indicate that when $n_c$ is equal to $n_b$ ($\alpha_{bc}=1$), optimal performance is achieved for a given memory limit, provided that $n_{ec}$ is sufficiently large. The number of extra neighboring chunks $n_{ec}$ and the number of rounds $r$ should be increased for larger batch sizes.
To demonstrate the effectiveness of our approach, we conducted a series of experiments in the quality-memory trade-off paradigm on a set of datasets (Section \ref{sec:datasets}), comparing the SASRec-RECE model with baselines described in Section \ref{sec:baselines}. In order to obtain quality metrics (NDCG@10) for different memory values, we evaluated these models for different values of hyperparameters affecting peak memory: batch size, number of negative samples for BCE$^+$, gBCE, and CE$^-$, $n_{ec}$ and $r$ for RECE. The grids for all parameters are available in our GitHub\footref{github}. Figure \ref{fig:baselines} shows optimal points for given memory budgets, dashed line means there was no configuration that showed higher quality at increased budget.

On BeerAdvocate, the dataset with a small catalog, SASRec-RECE achieves performance close to the best competitors (Fig. \ref{fig:baselines}a), both in terms of quality and peak memory. However, our method is more effective on datasets with larger catalogs, where the memory consumption problem becomes more pronounced. On Behance and Kindle Store datasets (Fig. \ref{fig:baselines}b-c), RECE provides nearly the same quality with $\textbf{12}$ (0.526GB vs. 6.48GB) and $\textbf{3}$ (3.15GB vs. 9.73GB) times less memory consumed, respectively, while on the Gowalla  (dataset with the largest catalog) (Fig. \ref{fig:baselines}d), RECE can either outperform the second best method by $\textbf{8.19\%}$ (0.0449 vs. 0.0415 NDCG@10) or yield almost the same quality with $\textbf{6.6}$ times less memory (0.78GB vs. 5.13GB). We can also see that SASRec-RECE does not produce any significant computational overhead, and its training time is at the lower end of the spectrum within the competitors (Fig. \ref{fig:baselines}e-h). Table \ref{tab:baselines} shows the extended set of metrics for the largest datasets.

\begin{table}[]
\setlength{\abovecaptionskip}{2pt} 
\caption{\textbf{{Performance under memory constraints. Bold indicates the best performance, and underlined is the second-best. NDCG@1 and HR@1 are equivalent by definition.}}} \label{tab:baselines}
\small
\begin{center}
\resizebox{1.\columnwidth}{!}{
\begin{tabular}{llccccc}
\hline
Dataset                                                                 & Model   & NDCG@1          & NDCG@5          & NDCG@10         & HR@5            & HR@10           \\ \hline
\multirow{5}{*}{\begin{tabular}[c]{@{}l@{}}Kindle\\ Store\end{tabular}} & BCE$^+$ & {\underline{0.0473}}    & \textbf{0.0674} & \textbf{0.0725} & \textbf{0.0856} & \textbf{0.1010} \\
                                                                        & gBCE    & {\underline{0.0473}}    & 0.0626          & 0.0681          & 0.0766          & 0.0933          \\
                                                                        & CE$^-$  & 0.0397          & 0.0582          & 0.0644          & 0.0744          & 0.0938          \\
                                                                        & CE      & 0.0467          & 0.0658          & 0.0711          & 0.0822          & 0.0972          \\
                                                                        & RECE    & \textbf{0.0486} & {\underline{0.0665}}    & {\underline{0.0714}}    & {\underline{0.0822}}    & {\underline{0.0985}}    \\ \hline
\multirow{5}{*}{Gowalla}                                                & BCE$^+$ & {\underline{0.0175}}    & {\underline{0.0341}}    & {\underline{0.0415}}    & {\underline{0.0507}}    & {\underline{0.0739}}    \\
                                                                        & gBCE    & 0.0165          & 0.0337          & 0.0410          & 0.0506          & 0.0733          \\
                                                                        & CE$^-$  & 0.0162          & 0.0321          & 0.0402          & 0.0475          & 0.0729          \\
                                                                        & CE      & 0.0164          & 0.0324          & 0.0388          & 0.0471          & 0.0674          \\
                                                                        & RECE    & \textbf{0.0192} & \textbf{0.0372} & \textbf{0.0449} & \textbf{0.0546} & \textbf{0.0787} \\ \hline
\end{tabular}
}
\end{center}
\end{table}

Finally, following \cite{Petrov_2023}, we evaluate SASRec-RECE on the Amazon Beauty dataset \cite{mcauley2015imagebased} against recent models, which report best results, including CBiT \cite{du2022contrastive}, DuoRec \cite{duo}, CL4SRec \cite{Xie2022ContrastiveLF}, and FEARec \cite{fear}. To comply with the evaluation protocols of these works, we replace the temporal data splitting strategy, described in Section \ref{sec:evals}, with the leave-one-out evaluation approach and the corresponding data preprocessing. Table \ref{tab:recentresults} shows that SASRec-RECE achieves performance comparable to the recently proposed models.

\begin{table}[]
\setlength{\abovecaptionskip}{4pt} 
\caption{\textbf{{Comparison of SASRec-RECE with recent results on Beauty using leave-one-out splitting strategy. Setup follows \cite{Petrov_2023}. Best values are in bold,  second best are underlined.}}} \label{tab:recentresults}
\resizebox{1\columnwidth}{!}{
\begin{tabular}{lccccc}
\hline
Metric & FEARec & CBiT & DuoRec & CL4SRec & SASRec-RECE \\ \hline
NDCG@10 & 0.0459 & \textbf{0.0537} & 0.0443 & 0.0299 & \underline{0.0525} \\
HR@10 & 0.0884 & \textbf{0.0905} & 0.0845 & 0.0681 & \underline{0.0897} \\ \hline
\end{tabular}%
}
\end{table}

\section{Conclusion}
\label{sec:conclusion}
In this work, we introduced RECE, a novel loss function that approximates full Cross-Entropy using hard-negative mining with GPU-efficient operations. This allows the benefits of CE loss, known for state-of-the-art performance, to be applied to large catalogs that would otherwise be infeasible due to high memory requirements. 
We demonstrated that RECE almost matches the performance and memory requirements of recent negative sampling methods on datasets with small catalogs and consumes up to $12$ times less memory on large-catalog datasets. Alternatively, it can improve quality (NDCG@10) by up to $8.19\%$ compared to other approaches if provided with an extended memory budget.
The idea behind RECE can potentially be applied not only to other loss functions and models in sequential recommender systems but also to other domains.

\begin{acks}
The work is supported in part through the Basic Research Program at the National Research University Higher School of Economics (HSE University).
\end{acks}

\bibliographystyle{ACM-Reference-Format}
\balance
\bibliography{content/8_bib}

\end{document}